%% file: Paper_7.tex
\documentclass[preprint,tightenlines,amsmath,amsfonts,amssymb,aps,showkeys]
              {revtex4}
\usepackage{latexsym,makeidx,ifthen,calc,longtable,graphics}
\usepackage{bm}

\begin{document}

\title{Measurement and Probability in Relativistic Quantum Mechanics}

\preprint{Version 4}

\author{Ed Seidewitz}
\email{seidewitz@mailaps.org}
\affiliation{235A E. Church St., Frederick MD 21701 USA}

\date{March 2025}


\begin{abstract}
Ultimately, any explanation of quantum measurement must be extendable
to relativistic quantum mechanics (RQM), since many precisely
confirmed experimental results follow from quantum field theory (QFT),
which is based on RQM. Certainly, the traditional ``collapse''
postulate for quantum measurement is problematic in a relativistic
context, at the very least because, as usually formulated, it violates
the relativity of simultaneity.  The present paper addresses this with
a relativistic model of measurement in which the state of the universe
is decomposed into decoherent histories of measurements recorded
within it.  The approach is essentially Everettian, in the sense that
it uses the unmodified, unitary quantum formalism of RQM. But it
addresses the difficulty with typical ``many worlds'' interpretations
on how to even define probabilities over different possible
``worlds''.  To do this, Zurek's concept of envariance is generalized
to the context of relativistic spacetime, giving an objective
definition of the probability of any one of the quantum histories,
consistent with Born's rule.  It is then shown that the statistics of
any repeated experiment within the universe also tend to follow the
Born rule as the number of repetitions increases.  The wave functions
that we actually use for such experiments are local reductions of very
coarse-grained superpositions of universal eigenstates, and their
``collapse'' can be re-interpreted as simply an update based on
additional incremental knowledge gained from a measurement about the
``real'' eigenstate of our universe.
\end{abstract}

\maketitle


\input{Paper_7_Body}

\bibliography{../../RQMbib}

\end{document}

%% file: Paper_7_Body.tex
\section{Introduction} \label{sect:intro}

The so-called \emph{measurement problem} has been discussed in the
literature of quantum theory for many years.  And one can safely say
that there is still no consensus on the solution for this ``problem''.
Nevertheless, the question is important for interpreting what we even
mean by experimental confirmation of quantum mechanics, and at the
foundations of the expanding technology of quantum computing.

The traditional simple model for a quantum measurement considers a
quantum system $\sysS$ that is measured by a macroscopic apparatus
$\sysA$, interacting over time according to Schr\"odinger's equation.
$\sysS$ and $\sysA$ begin in initial states $\ket{s_{0}}$ and
$\ket{a_{0}}$, respectively, and then evolve over some period of time
into a joint entangled state, such that the so-called \emph{pointer
states} $\ket{a_{i}}$ of $\sysA$ become correlated with the measurable
eigenstates $\ket{s_{i}}$ of $\sysA$:
\begin{equation*}
	\ket{s_{0}}\ket{a_{0}} \rightarrow
		\sum_{i} \psi_{i}\ket{s_{i}}\ket{a_{i}} \,.
\end{equation*}
The probability of obtaining measurement result $i$ is then given by 
$\sqr{\psi_{i}}$, following the Born rule \cite{born26}.

Of course, actual measurement processes result in specific measurement
results.  The dynamic theory, however, does not determine which result
is actually chosen, only the probabilities.  The effect of a
measurement process discontinuously \emph{projecting} the entangled
state onto a specific eigenstate $\ket{s_{i}}\ket{a_{i}}$ has to be
added as an additional postulate \cite{vonneumann55}.  

This issue of \emph{definite outcomes} is what is typically called the
``measurement problem'' in quantum mechanics.  But there are other
important related issues that have also been well-discussed in the
literature over the years.  These include, in particular, the
following two.

\begin{itemize}
	\item \emph{What actually constitutes a ``measurement process'' such
	that it causes a ``collapse'' of the system state?}
	
	It seems to have become generally accepted that interaction with
	the macroscopic environment results in the \emph{decoherence} of a
	measurement superposition state into a mixed state with
	classical probabilities $\sqr{\psi_{i}}$
	\cite{gellmann90,griffiths84,omnes88}.  But what is the role of a
	measurement apparatus (or human observer), if any, in causing this
	decoherence and specifically amplifying and recording a system
	measurement?

	\item \emph{What even is the meaning of ``probability'' in the
	interpretation of the theory for actual, physical measurement
	processes?  And why the odd Born rule for the probabilities?}
	
	A common approach is to define probabilities based on the relative
	frequencies of each of the possible measurement outcomes as the
	same measurement experiment is carried out repeatedly (for
	example, \refcites{everett57,hartle68,graham73}).  However using
	this approach to derive the Born rule has often been criticized as
	circular or inconsistent in the end (see, for example,
	\refcites{kent90,squires90,zurek05}).  In particular, the
	formalism of decohered reduced density matrices already
	presupposes the Born rule.
	
	A more recent alternative has been to apply Bayes' Theorem
	\cite{bayes63}, so that ``probability quantifies a degree of
	belief for a single trial, without any a priori connection to
	limiting frequencies'' \cite{caves02,fuchs10}.  The cost,
	however, is that such a definition of probability is inherently
	\emph{subjective}, relying on an understanding of how the ``level
	of ignorance'' or ``degree of belief'' of a rational agent should
	be adjusted based on observation.

\end{itemize}

Now, bringing relativistic quantum mechanics (RQM) into this
discussion would only seem to make it more difficult.  The very
concepts of dynamic evolution over time and ``spontaneous collapse''
cannot easily be made relativistically Lorentz covariant.  But we know
the world is relativistic, and many of the most accurate predictions
of quantum mechanics are fundamentally based on relativistic quantum
field theory (QFT).  So any solution to the measurement problem is
going to have to somehow be in the context of RQM
\cite{wallace22,adlam23}.

Attempts at dealing mathematically with measurements in RQM often use
specific non-collapse formalisms, such as decohering
spacetime path integrals \cite{hartle95,hartle97,halliwell92,%
halliwell98,halliwell01a,halliwell03,halliwell04,seidewitz06a,%
seidewitz06b,seidewitz09,seidewitz11} or relativistic Bohmian
mechanics \cite{bohm52,bohm93,nikolic07,nikolic12}.  As a result,
these approaches are not widely familiar and not always easily
connected with the typical formalisms of QFT. However, in this paper I
will show that it is instead possible to address the issues of
measurement and probability relativistically using only a basic,
traditional formalism for RQM.

Since the approach uses the formalism of unitary RQM, it is
essentially Everettian \cite{everett57,dewitt73,everett73}, and yet it
dispenses with the usual conception of a ``branching'' of a system
state as it is measured over time.  Instead, the manifestly covariant
spacetime formalism of RQM leads to the decomposition of the entire
``state of the universe'' into orthogonal eigenstates distinguished by
having different sets of mutually exclusive measurement results.  Each
of these eigenstates represents an entire, distinct, possible history
of the universe over all spacetime.  

Quantum theory then gives a probability for each one of these
histories that it is the history of our actual universe.  From
``inside'' the actual universe, our experiments provide us information
on what the state of the universe actually is, and the statistical
probability of results is determinable from the predicted probability
of the state of the universe as a whole, contingent on our prior
knowledge of that state.  Subsequent sections of this paper formalize
this insight, showing why the observed statistical distribution of
quantum measurement results follows Born's rule, without recourse to
any ``collapse'' postulate, consistent with RQM.

\Sect{sect:formalism} begins with an outline of the traditional
formalism of Hilbert space and projection operators, in the context of
RQM, and introduces the concept of the ``state of the universe''.
\Sect{sect:measurements} then addresses the issue of how to model
measurements using this formalism.  \Sect{sect:prob} tackles the issue
of providing a classical probability interpretation for such
measurements, and \sect{sect:stats} applies this to predicting the
statistical results of repeated experiments.
\Sect{sect:interpretation} discusses the interpretational implications
of the results presented in the previous sections, and
\sect{sect:conclusion} gives some concluding philosophical remarks.

\section{Relativistic Quantum Mechanics} \label{sect:formalism}

We begin with the Hilbert space formalism common to both
non-relativistic and relativistic quantum mechanics.  As usual, a
quantum \emph{state} $\ketPsi$ is defined as a normalizable vector in
a separable Hilbert space $\HilbH$.  We will typically take states to
already be normalized to $\sqr{\Psi} = 1$.  An \emph{operator} $\opA$
then maps a state $\ketPsi \in \HilbH$ to another state $\opA\ketPsi$.

An \emph{observable} quantity is represented as a self-adjoint
operator on $\HilbH$.  This paper is concerned with the
\emph{measurement} of such observables.  For that purpose, it will be
sufficient to assume an observable $\opA$ has a discrete spectrum of
real eigenvalues $a_{i}$ with a corresponding set of normalized
eigenstates $\ket{a_{i}}$.

(Of course, there are important observables, such as position and
velocity, that are generally modeled as having continuous spectra.
Nevertheless, any real measurement of such continuous observables will
be discrete, limited by the accuracy of the measurement instrument.
Thus, measurement effectively discretizes a continuous observable,
integrating over finite intervals of its spectrum.  Assuming this will
be sufficient for our considerations in \sect{sect:measurements}.)

Moving now to RQM, we need to define what it means to be
relativistically invariant under an arbitrary \emph{Poincar\'e
transformation} $\{\Delta x, \Lambda\}$, where $\Delta x$ is a
(4-dimensional) spacetime translation and $\Lambda$ is a Lorentz
transformation.  To do this, there must be a unitary operator
representation of the Poincar\'e group defined on $\HilbH$.  That is,
for any Poincar\'e transformation $\{\Delta x, \Lambda\}$, there
exists a unitary operator $\opU(\Delta x, \Lambda)$ that has the
effect of applying the transformation to states in $\HilbH$,
satisfying the group property for composing such transformations. An 
operator can then be considered relativistically invariant if it 
commutes with $\opU$ for all possible transformations.

As in non-relativistic quantum mechanics, the dynamics of the theory
are given by a self-adjoint \emph{Hamiltonian operator} $\opH$, which
is required to commute with $\opU$, so that it is relativistically
invariant under Poincar\'e transformations.  But this means that the
Hamiltonian can no longer be considered to be simply a time evolution
operator on states, because coordinate time evolution is not
relativistically invariant.  Instead, the Hamiltonian acts as a
constraint on which states in $\HilbH$ are considered \emph{physical}.
Specifically, a \emph{physical state} is one that satisfies
$\opH\ketPsi = 0$.  This forms a subspace $\HilbH\phys$ of $\HilbH$
containing all physical states meeting this constraint.

For a free theory, $\opH = \opP^{2} + m^{2}$ (where $\opP$ is the
momentum operator defined by $\opU(\Delta x, 0) =
\exp(-i\opP\cdot\Delta x)$), essentially enforcing an on-shell mass
condition.  For an interacting theory, $\opH$ introduces all the
dynamics of the theory.  Particularly for a theory that includes
gravity, the constraint $\opH\ketPsi = 0$ is known as the
Wheeler-DeWitt equation \cite{dewitt67}.  (See also
\refcites{dirac50,dirac64} on Dirac's theory of constraints in
Hamiltonian mechanics and \refcite{rovelli04} on Hamiltonian
constraint mechanics and solving the Wheeler-DeWitt equation in the
context of quantum gravity.)

From now on, we will only consider states in the physical subspace 
$\HilbH\phys$ and require that all operators map states in 
$\HilbH\phys$ into states in $\HilbH\phys$. That is, all physical 
operators must commute with the Hamiltonian $\opH$. For simplicity, 
we will now also drop the subscript ${}_\text{phys}$ and just write 
$\HilbH$ for the physical subspace, since that is all that will 
concern us going forward.

This paradigm for ``dynamics'' is clearly quite different than in
non-relativistic dynamics.  Since all states and operators are
required to be stationary with respect to the Hamiltonian, we can no
longer treat a system as ``starting'' in a certain initial state and
``evolving'' under the Hamiltonian over time.  We need a different
vocabulary to discuss how ``dynamics'' happens in a manifestly
Poincar\'e-invariant spacetime framework.

As discussed above, the condition $\opH\ketPsi = 0$ constrains a
physical state $\ketPsi$ effectively ``over all spacetime''. However, 
observables of such a state can be parameterized by spacetime points. 
And, in the coordinate system of a specific observer, such 
observables will seem to ``evolve'' forward in coordinate time, 
consistent with the dynamic constraint imposed by $\opH$ on $\ketPsi$.

Nevertheless, for the purposes of this paper, we would like to both
abstract away from the details of the Hamiltonian and remain
manifestly independent of any particular coordinate system.  To do
this, we first decompose any observable $\opA$ (with a discrete
spectrum) as
\begin{equation*} \label{eqn:B1}
	\opA = \sum_{i} a_{i} \opP_{A}^{i} \,,
\end{equation*}
where the $a_{i}$ are its eigenvalues and the $\opP_{A}^{i}$ are
\emph{projection operators} defined by
\begin{equation*} \label{eqn:B2}
	\opP_{A}^{i} \equiv \ket{a_{i}}\bra{a_{i}} \,.
\end{equation*}
A projection operator $\opP$ characterizes a subspace $\opP\HilbH$ of
$\HilbH$ such that $\opP\ketPsi \in \opP\HilbH$ for any $\ketPsi \in
\HilbH$.  Such a projection operator can then be interpreted as
representing a logical \emph{proposition} that is true for the states
$\ketPsi \in \opP\HilbH$ (i.e., for which $\ketPsi$ is a unit
eigenvector of $\opP$), false for the states for which $\opP\ketPsi =
0$ and undetermined otherwise \cite{vonneumann55}.  We can also say
that $\opP$ describes a \emph{condition} that holds
for those $\ketPsi \in \opP\HilbH$.

In particular, suppose $\{\opP_{\sysS}^{\alpha}\}$ is a set of projection
operators representing conditions characterizing a system $\sysS$, 
with each $\opP_{\sysS}^{\alpha}$ representing the proposition that
system $\sysS$ is in some condition $\alpha$. We can require that 
these conditions be \emph{independent}, meaning that the subspaces so 
identified for each $\alpha$ must be mutually orthogonal:
\begin{equation*} \label{eqn:B3}
	\opP_{\sysS}^{\alpha}\opP_{\sysS}^{\beta}
		= \delta_{\alpha\beta}\opP_{\sysS}^{\alpha} \,.
\end{equation*}
And we can further require that the conditions be \emph{complete},
meaning that the system must be in one of the possible conditions for
it:
\begin{equation*} \label{eqn:B4}
	\sum_{\alpha} \opP_{\sysS}^{\alpha} = 1 \,.
\end{equation*}

There can be multiple such complete, independent sets of projection
operators giving different characterizations of a system.  For
example, suppose the set $\{\opP_{\sysS}^{\alpha}(x)\}$ represents
conditions of system $\sysS$ at some spacetime point $x$ (or, for a
physically extended system, some spacetime volume around $x$).  Then,
for a specific system state $\ketPsi_{\sysS}$, the system can be, say,
in both condition $\alpha$ at $x_{1}$ and condition $\beta$ at $x_{2}$
if and only if $\opP_{\sysS}^{\beta}(x_{2})
\opP_{\sysS}^{(0)\alpha}(x_{1}) \ketPsi_{\sysS}$ is non-zero.
Further, if $x_{2}$ is timelike separated from and (invariantly) later
than $x_{1}$, this effectively determines whether the Hamiltonian
dynamics allows the system to ``evolve'' from $x_{1}$ to $x_{2}$.

We can also consider separate conditions of \emph{multiple} subsystems
of an overall physical state.  For example, consider two systems
$\sysS_{1}$ and $\sysS_{2}$ with corresponding conditions
$\{\opP_{\sysS_{1}}^{\alpha_{1}}\}$ and
$\{\opP_{\sysS_{2}}^{\alpha_{2}}\}$.  These operators operate on the
\emph{same} combined physical state $\ketPsi_{\sysS_{1}\sysS_{2}}$, so
it is possible that the Hamiltonian constraint
$\opH\ketPsi_{\sysS_{1}\sysS_{2}} = 0$ causes the two systems to
become \emph{correlated} such that
\begin{equation*}
	\opP_{\sysS_{2}}^{\alpha_{2}}
	\opP_{\sysS_{1}}^{\alpha_{1}}\ketPsi_{\sysS_{1}\sysS_{2}}
		= \delta_{\alpha_{2},A(\alpha_{1})}
		  \opP_{\sysS_{1}}^{\alpha_{1}}
		  \ketPsi_{\sysS_{1}\sysS_{2}} \,.
\end{equation*}
This means that, if $\sysS_{1}$ is in some condition $\alpha_{1}$,
then we can expect $\sysS_{2}$ to be in condition $\alpha_{2} =
A(\alpha_{1})$, where $A$ is some discrete function from conditions of
$\sysS_{1}$ to corresponding conditions of $\sysS_{2}$.  We can treat
this correlation as the result of the \emph{interaction} between the
systems under the constrained Hamiltonian dynamics.  This conception
of ``interaction'' will be critical in our analysis of measurement
processes in \sect{sect:measurements}.  However, first we need to take
one more step with our formalism for RQM.

As reiterated by Zurek, ``The Universe consists of systems''
\cite{zurek07b}.  Quantum processes, measuring apparatuses and
observers are all systems, and all subsystems of the system that is
the universe as a whole.  Further, particularly when considering
decoherence and records of measurements, one must consider that there
is always some interaction with the ``rest of the universe''.  It is
therefore convenient to consider physical states in $\HilbH$ to simply
be states of \emph{the entire universe} over all space and time,
characterizing any system \emph{within} the universe relative to a
specific overall state \emph{of} the universe.

The concept of a ``state of the universe'' actually goes back at least
to Everett's original formulation of the \emph{universal wave
function} (``state function of the whole universe'') \cite{everett73},
though this explicit concept was removed from his thesis as first
published \cite{everett57}.  Hartle and Hawking \cite{hartle83} later
proposed a ground-state \emph{wave function of the universe} that
gives ``the amplitude for the Universe to appear from nothing''.
Hartle and Hawking make no reference to the earlier work of Everett in
their paper, because they were dealing with technical issues of how to
formulate such a wave function for quantum gravity, not questions of
interpretation.  But further work of Hartle and colleagues can be seen
exactly as addressing such questions \refcites{hartle95,hartle97,%
halliwell92,halliwell98,halliwell01a,halliwell03,halliwell04}.

Of course, given a state of the universe, one still needs to have some
way to discuss the interaction of systems \emph{within} the universe,
where a \emph{system} is any part of the universe that can be
delineated from the rest of the universe.  But we can do this using
the formalism that has already been discussed in this section.  For
any specific system, some propositions about the universe as a whole
will concern the condition of that system, while some will not.  And
the Hamiltonian constraint on the overall state of the universe
determines correlations between the conditions pertaining to different
systems.

So, we will characterize a system $\sysS$ system by a set of
(independent and complete) propositions $\{\opP_{\sysS}^{\alpha}\}$
that can be made concerning it, which essentially defines what we are
interested in ``knowing'' about it.  It should be kept in mind that
such operators now act on the state of the universe, and therefore
define propositions on \emph{all of spacetime} in that state.  Some
care must be taken in considering the completeness of the propositions
associated with a system, in that there will be many states of the
universe in which the system will essentially \emph{not exist}.

For example, suppose that the system of interest is a measuring
instrument and that the projection operators characterizing it
represent pointer states of the instrument.  But this presumes that
the instrument is \emph{actually there} (and turned on, and operating,
etc.).  There will be many states of the universe in which this is
simply not the case.

By convention, for any system $\sysS$, take the condition $\alpha = 
0$ to represent the system ``not existing''. The actually interesting 
conditions of the system are indexed by $\alpha > 0$. By definition, 
the non-existence assertion $\opP_{\sysS}^{0}$ is never truly 
interesting (at least for the cases considered here), but it is only 
when this operator is included that the set of 
$\opP_{\sysS}^{\alpha}$ is complete.

And this now completes the RQM formalism needed to proceed with our 
analysis of measurement processes and probabilities.

\section{Measurements and Records} \label{sect:measurements}

In general, the actual dynamics of the theory will not be important
for this paper.  However, as discussed in \sect{sect:formalism}, due
to the constraint $\opH\ketPsi = 0$, two systems $\sysA$ and $\sysS$
may interact such that the condition of $\sysA$ is fully or partially
determined by the condition of $\sysS$.  In particular, if $\sysA$ is
intended as a \emph{measuring apparatus}, then its conditions will be
adjusted so that they are perfectly correlated with the conditions of
interest of $\sysS$, with $\alpha$ being the pointer conditions
corresponding to the measured quantities $\beta$:
\begin{equation} \label{eqn:C1}
	\opP_{\sysA}^{\alpha}\opP_{\sysS}^{\beta}\ketPsi
		= \delta_{\alpha\beta}\opP_{\sysS}^{\beta}\ketPsi \,.
\end{equation}
(Note that, in this case, $\sysA$ might actually ``exist'' even in the
condition $\alpha = 0$ that corresponds to $\beta = 0$, in which
$\sysS$ does not exist.  But this is still an uninteresting condition,
since no measurement can be made anyway when $\sysS$ does not exist.)

For a proper measurement, simple correlation is not enough, though.
$\sysA$ must also leave a \emph{record} of its result.  This record is
left through interaction of $\sysA$ with the \emph{environment}
$\sysE$, which is the rest of the universe other than $\sysA$ and
$\sysS$.  And this record must be independent of any interaction of
the environment with $\sysS$.  (Indeed, to ensure classicality, the
record should be redundant and accessible to observers, though these
details will not be critical here.  This does require that the
environment have significantly more degrees of freedom than either the
system being measured or the measurement apparatus, but that is always
the case for any real measurement.  See also
\refcites{zurek09,riedel16,zurek22}.)

In effect, the environment is another (ever present) system whose
conditions of interest (in this case) are the records correlated with
the measurement results of $\sysA$:
\begin{equation} \label{eqn:C2}
	\opP_{\sysE}^{\alpha}\opP_{\sysA}^{\beta}\ketPsi
		= \delta_{\alpha\beta}\opP_{\sysA}^{\beta}
		  \ketPsi \,.
\end{equation}
Using the completeness of the $\opP_{\sysA}^{\gamma}$ with 
\eqns{eqn:C1} and \eqref{eqn:C2} then gives:
\begin{equation} \label{eqn:C3}
	\begin{split}
	\opP_{\sysE}^{\alpha}\opP_{\sysS}^{\beta}\ketPsi
		&= \sum_{\gamma} \opP_{\sysE}^{\alpha}
		   \opP_{\sysA}^{\gamma}\opP_{\sysS}^{\beta}\ketPsi \\
		&= \sum_{\gamma} \delta_{\alpha\gamma}\delta_{\gamma\beta} 
		   \opP_{\sysS}^{\beta}\ketPsi \\
		&= \delta_{\alpha\beta}\opP_{\sysS}^{\beta}\ketPsi\,.
	\end{split}
\end{equation}

It is not particularly important here what basis is chosen for the
measurement in \eqn{eqn:C1}.  However, it is the three-way correlation
between the environment, the apparatus and the system that ensures
that the chosen basis is unambiguous and physical for the recorded
measurement \cite{elby94}.  For the rest of the discussion here,
though, it will be sufficient to assume the correlation between the
condition of a system and a record of its condition in the
environment, as in \eqn{eqn:C3}, without explicitly including the
measuring apparatus that led to that record.  (For more on the issue
of basis selection, in a non-relativistic context, see the description
of \emph{einselection} and \emph{quantum Darwinism} by Zurek et al.
\cite{zurek82,zurek98,zurek03a,zurek07b,zurek09,riedel16,zurek22}.)

So, consider that
\begin{equation*} \label{eqn:C4}
	\opP_{\sysS}^{\beta}\ketPsi
		= \psi_{\sysS}^{\beta}(\Psi)\ket{s^{\beta}(\Psi)} \,,
\end{equation*}
where $\ket{s^{\beta}(\Psi)}$ is a unit eigenstate of
$\opP_{\sysS}^{\beta}$ and $\psi_{\sysS}^{\beta}(\Psi)$ is the
magnitude of $\opP_{\sysS}^{\beta}\ketPsi$.  Further, since the record
in the environment is independent of interaction with $\sysS$, we can
take $\opP_{\sysE}^{\alpha}$ and $\opP_{\sysS}^{\beta}$ to commute, so
$\ket{s^{\beta}(\Psi)}$ is actually a joint eigenstate of
$\opP_{\sysE}^{\alpha}$ and $\opP_{\sysS}^{\beta}$.  Therefore:
\begin{equation*} \label{eqn:C5}
	\opP_{\sysE}^{\alpha}\opP_{\sysS}^{\beta}\ketPsi
		= \delta_{\alpha\beta}\psi_{\sysS}^{\beta}(\Psi)
		  \ket{e^{\alpha}(\Psi)s^{\beta}(\Psi)} \,.
\end{equation*}
Since the $\opP_{\sysE}^{\alpha}$ and $\opP_{\sysS}^{\beta}$ are 
complete sets,
\begin{equation} \label{eqn:C6}
	\begin{split}
	\ketPsi &= (\sum_{\alpha}\opP_{\sysE}^{\alpha})
			   (\sum_{\beta}\opP_{\sysS}^{\beta})\ketPsi \\
			&= \sum_{\alpha\beta}\opP_{\sysE}^{\alpha}
			   \opP_{\sysS}^{\beta}\ketPsi \\
			&= \sum_{\alpha}\psi_{\sysS}^{\alpha}
			   \ket{e^{\alpha}s^{\alpha}}
	\end{split}
\end{equation}
(where the notation has been simplified by eliding explicit
functional dependence on the state $\Psi$).  

Because of the required orthogonality of the set of projection
operators for a system, \eqn{eqn:C6} is a complete orthogonal
decomposition of the state $\ketPsi$.  Each of the eigenstates
$\ket{e^{\alpha}s^{\alpha}}$ represents a \emph{branch} of the
universe in which $\sysE$ records that $\sysS$ is in condition
$\alpha$.  This is similar to the decomposition into branch states in
consistent or decoherent histories formalisms
\cite{griffiths84,omnes88,gellmann90,griffiths02}, but, rather than
representing a time-ordered history of quantum propositions, the
branch states here are eigenstates of the universe \emph{for all
spacetime}.

If we were to now apply Born's rule directly to \eqn{eqn:C6}, then we
would interpret the squared amplitudes $\sqr{\psi_{\sysS}^{\alpha}}$
as the probabilities for the universe to be in the branch
$\ket{e^{\alpha}s^{\alpha}}$---that is, for the system $\sysS$ to be
in the condition $\alpha$.  But how do we understand what a
probability even \emph{is} when we are considering different branches
of the entire universe?  We turn to this issue next.

\section{Probability} \label{sect:prob}

Given the decomposition in \eqn{eqn:C6}, the universe can ``really''
only be in one of the states $\ket{e^{\alpha}s^{\alpha}}$, because
each of these states asserts a proposition about the system $\sysS$
that is mutually incompatible with the proposition asserted by the
other states.  We therefore interpret the weighted superposition of
these states in $\ketPsi$ as meaning that they are all ``possible'',
and we would like the theory to tell us their relative
\emph{likelihood}.  This interpretation is, of course, a presumption,
but it is a fundamental postulate of quantum theory that seems to
accurately represent our reality.

It is well-known that Gleason's theorem requires the Born rule when
defining an additive probability measure on a Hilbert space (of
dimension at least three) \cite{gleason57}.  And Hossenfelder has
recently shown the rule can be derived even more simply assuming only
a distribution that is ``continuous, independent of [dimension], and
invariant under unitary operations'' \cite{hossenfelder21}.  But, as
Zurek remarks, mathematical derivations such as these give ``no
physical insight into why the result should be regarded as
probability'' \cite{zurek07b}.

So, what \emph{is} ``probability''?  This is a much discussed
philosophical question.  For our purposes, though, we can consider a
simple conception essentially underlying both the frequentist and
Bayesian approaches to probability, as stated by Laplace (though
Bayes' work actually preceded Laplace's publication): probability is
``the ratio of the number of favorable cases to that of all the cases
possible'' \cite{laplace51}, and the ``cases favorable to the event
being sought'' are considered to be all equally likely.  This leads to
a \emph{principle of indifference}, such that, if a change to the
system under consideration simply moves it from one ``favorable case''
to another, then an observer will be indifferent to such a change in
computing probabilities.

In this regard, Zurek notes that, if we can define a physical system
symmetry among ``favorable cases'', then we could make the subjective
principle of indifference into a definition of \emph{objective}
probabilities.  He proposes \emph{entanglement-assisted invariance}
(``envariance'') as the symmetry to do this
\cite{zurek82,zurek03a,zurek03b,zurek05,zurek07b}.  Envariance is a
symmetry of an entangled state in which an operation on one of the two
entangled systems can be undone by an operation solely on the other.
Zurek argues that if this leaves the overall entangled state
invariant, then we must consider all the various decompositions of the
entangled state related by this symmetry to be physically
equivalent---that is, we are effectively ``ignorant'' as to which is
the ``real'' decomposition.

By intent, this symmetry is tailored for the situation resulting from
a quantum measurement, as in \eqn{eqn:C6}, which is an entanglement
between the system $\sysS$ and its environment $\sysE$. Zurek uses 
envariance to analyze this situation in two steps.

\begin{enumerate}

	\item Add an arbitrary phase $\sigma_{\alpha}$ to each of the
	projection operators $\opP^{\alpha}_{\sysS}$ for the system $\sysS$:
	\begin{equation*}
		\opP^{\alpha}_{\sysS} \rightarrow 
			\me^{\mi \sigma_{\alpha}} \opP^{\alpha}_{\sysS} \,.
	\end{equation*}	
	This operation acts solely on the subspaces of $\HilbH$ corresponding
	to the conditions of $\sysS$, resulting in a new state of the universe
	$\ket{\Psi'}$ with the same branch states, but changes in the phases
	of the coefficients $\psi_{\sysS}^{\alpha}$.

	Now make a similar, but inverse, operation on the projection
	operators $\opP^{\alpha}_{\sysE}$ for the environment $\sysE$:
	\begin{equation*} 
		\opP^{\alpha}_{\sysE} \rightarrow 
			\me^{-\mi\sigma_{\alpha}} \opP^{\alpha}_{\sysE} \,.
	\end{equation*}	
	This operation acts solely on the subspaces of $\HilbH$ corresponding
	to the conditions of $\sysE$.  But, because of the entanglement of
	$\sysS$ and $\sysE$, it results in a complete reversal of the
	operation on $\sysS$, changing $\ket{\Psi'}$ back into $\ketPsi$.
		
	Thus, the state $\ketPsi$ as given in \eqn{eqn:C6} is envariant under
	the combined action of the above operations on $\sysS$ and $\sysE$.
	That is, the effect on the universe of an operation on the system
	$\sysS$ can be undone by an operation on the environment
	alone---effectively just a compensating adjustment in how the
	condition of the system is recorded.  As argued by Zurek, this implies
	that the original operation on $\sysS$ cannot be considered physically
	significant, which means that the phases of the coefficients
	$\psi_{\sysS}^{\alpha}$ cannot have physical significance.
		
	\item Swap two specific conditions $\beta \leftrightarrow \gamma$ in
	$\sysS$ (for $\beta \neq \gamma$):
	\begin{equation*}
		\opP^{\beta}_{\sysS} \leftrightarrow \opP^{\gamma}_{\sysS} \,.
	\end{equation*}
	Then also swap the corresponding conditions of $\sysE$:
	\begin{equation*}
		\opP^{\beta}_{\sysE} \leftrightarrow \opP^{\gamma}_{\sysE} \,.
	\end{equation*}
	The result of applying both these operations to $\ketPsi$ is a new
	state in which the coefficients are switched on the $\beta$ and
	$\gamma$ terms:
	\begin{equation*}
		\ket{\Psi'} = 
			\psi_{\sysS}^{\beta}\ket{e^{\gamma},s^{\gamma}}
			+ \psi_{\sysS}^{\gamma}\ket{e^{\beta},s^{\beta}}
			+ \ldots \,.
	\end{equation*}

	Clearly, this will not, in general, be the same as $\ketPsi$.
	However, if it happens that $\psi_{\sysS}^{\beta} =
	\psi_{\sysS}^{\gamma}$, then, in fact, $\ket{\Psi'} = \ketPsi$.  That 
	is, in this case, the action of the swap operation on $\sysS$ is 
	undone by the swap operation on $\sysE$. Indeed, this case 
	corresponds to essentially just relabelling the states for the 
	$\beta$ and $\gamma$ conditions, which certainly shouldn't be 
	physically relevant. 
\end{enumerate}

So, this establishes the desired physical symmetry for defining
objective probabilities: states that are physically indistinguishable
when swapped should be considered equally likely.  In particular,
suppose all the coefficients in \eqn{eqn:C6} have equal magnitude.
Then all the branch states represent equally likely branches of the
universe, since swapping any one of them with any other leaves the
overall state of the universe unchanged, with the swapped conditions
simply relabelled.  Indeed, the normalization normalization
$\sum_{\alpha} \sqr{\psi_{\sysS}^{\alpha}} = 1$ implies that, if all
the $\psi_{\sysS}^{\alpha}$ have equal magnitude, then
$|\psi_{\sysS}^{\alpha}| = \sqrt{1/N}$, where $N$ is the number of
$\alpha$ conditions, and
\begin{equation*}
	\mathrm{Prob}(\ket{e^{\alpha},s^{\alpha}}) 
		= \frac{1}{N}
		= \sqr{\psi_{\sysS}^{\alpha}} \,,
\end{equation*}
which is just the Born rule for this case.

But, of course, the coefficients in \eqn{eqn:C6} will generally
\emph{not} have equal magnitude.  We can address this by breaking down
each condition $\alpha$ of $\sysS$ into a set of finer-grained
conditions, such that the states for these fine-grained conditions
\emph{do} have equal magnitude.  We do this by introducing an
ancillary system $\sysC$ with doubly-indexed projection operators
$\opP_{\sysC}^{\alpha\beta}$ such that
\begin{equation} \label{eqn:E1}
	\opP_{\sysC}^{\alpha\beta} \ket{e^{\alpha},s^{\alpha}}
		= \ket{c^{\alpha\beta},e^{\alpha},s^{\alpha}} /
		  \sqrt{m_{\alpha}} \,, \quad \beta = 1,\dots, m_{\alpha} \,,
\end{equation}
for unit eigenstates $\ket{c^{\alpha\beta},e^{\alpha},s^{\alpha}}$ of
$\opP_{\sysC}^{\alpha\beta}$, where the coefficients
$1/\sqrt{m_{\alpha}}$ are chosen to preserve the normalization of
$\ket{e^{\alpha},s^{\alpha}}$.

Now, mathematically, it is always possible to introduce an ancillary
system such as $\sysC$ to expand the Hilbert space, by the Stinespring
dilation theorem \cite{stinespring55}.  However, considering $\sysC$
as a physical system, it is essentially a finer-grained measuring
apparatus, with a subdivision of the environment to allow a
finer-grained recording of the condition of $\sysS$.  Constructing
such a system is also always possible, at least in principle, because
of the vastly larger number of degrees of freedom already required in
the environment to keep proper records.  As noted by Zurek, ``finding
$\sysC$ with the desired dimensionality of the respective subspaces
and correlating it with $\sysS$ in the right way is not a
`hit-or-miss' proposition---it can always be accomplished using the
information in the observer's possession.  It is also straightforward
in principle to find the environmental degrees of freedom that would
decohere the fine-grained states of $\sysC$\ldots'' (from Sect.\ V.C
of \refcite{zurek05}).

Inserting \eqn{eqn:E1} into \eqn{eqn:C6} gives:
\begin{equation*}
	\ketPsi = \sum_{\alpha\beta} 
		\frac{\psi^{\alpha}_{\sysS}}{\sqrt{m_{\alpha}}}
		\ket{c^{\alpha\beta},e^{\alpha},s^{\alpha}} \,,
\end{equation*}
and we want to make all the coefficients equal:
\begin{equation*}
	\frac{\psi^{\alpha}_{\sysS}}{\sqrt{m_{\alpha}}}
		= \psi_{\sysS\sysC} \,, \text{ for all $\alpha$.}
\end{equation*}
The normalization $\sum_{\alpha} \sqr{\psi_{\sysS}^{\alpha}} = 1$ then
implies $ \psi_{\sysS\sysC} = 1 / \sqrt{M}$, where $M = \sum_{\alpha}
m_{\alpha}$, giving
\begin{equation*}
	\ketPsi = \sum_{\alpha\beta} \sqrt{\frac{1}{M}}
			  \ket{c^{\alpha\beta},e^{\alpha},s^{\alpha}} \,.
\end{equation*}

The coefficients in this decomposition are now, by construction, all
the same.  Therefore, as previously argued, each of the states
$\ket{c^{\alpha\beta},e^{\alpha},s^{\alpha}}$ can be considered to be
equally likely.  Since there are a total of $M =
\sum_{\alpha}m_{\alpha}$ terms,
\begin{equation*} \label{eqn:E3}
	\mathrm{Prob}(\ket{c^{\alpha\beta},e^{\alpha},s^{\alpha}})
		= \frac{1}{M}\,.
\end{equation*}
For each $\alpha$, $m_{\alpha}$ of the overall
system/en\-vi\-ron\-ment/an\-cilla states correspond to the system
outcome $\alpha$, so the probability for this outcome is
\begin{equation*} \label{eqn:E4}
	\mathrm{Prob}(\ket{e^{\alpha},s^{\alpha}})
		= m_{\alpha} \mathrm{Prob}(
			\ket{c^{\alpha\beta},e^{\alpha},s^{\alpha}})
		= \frac{m_{\alpha}}{M} = \sqr{\psi_{\sysS}^{\alpha}} \,,
\end{equation*}
which is just the Born rule.  (This assumes the additivity of
probabilities, but it is possible to come to the same conclusion
without making this assumption \cite{zurek05}.)  Now, the above
clearly requires that $\sqr{\psi^{\alpha}_{\sysS}}$ is a rational
number, since $m_{\alpha}$ and $M$ are both natural numbers.
Nevertheless, the derivation can be extended to to irrational
$\sqr{\psi_{\sysS}^{\alpha}}$ by continuity.

(Zurek also addresses``Born's Rule for Continuous Spectra'' in an
appendix to \refcite{zurek05}, in a non-relativistic context.
Providing a relativistic generalization of this will be considered in
future work.  However, as previously noted, this should not
fundamentally change the conclusions of the present work.)

Of course, this only establishes the probability interpretation for
the coefficients of branch states such as
$\ket{e^{\alpha},s^{\alpha}}$ in the expansion of the state of the
universe.  It is not immediately clear how such an assignment of
probabilities, essentially for a state of the entire universe, relates
to the statistics of the physical results of measurement processes
occurring within the universe.  To clearly establish this
relationship, let us consider how such statistics are computed.

\section{Statistics} \label{sect:stats}

Suppose the same experiment is repeated, independently, $n$ times.
The ``same experiment'' means identical experimental setups, but each
run of the experiment is still a separate system in the universe, as
defined in \sect{sect:formalism}.  Let $\sysS_{i}$, for $i =
1,\dots,n$, be the systems representing the repeated runs of the
experiment, giving the results $m_{i}$.

Since the systems are independent, we can define 
joint eigenstates of the projection operators for the experimental 
results for each of the systems:
\begin{equation*} \label{eqn:E5}
	\opP_{\sysS_{i}}^{m}\ket{m_{1}, \dots, m_{n}}
		= \delta_{mm_{i}}\ket{m_{1}, \dots, m_{n}} \,.
\end{equation*}
Keeping the common environment of all the systems implicit, we can
then decompose the state of the universe $\ketPsi$ as
\begin{equation} \label{eqn:E6}
	\ketPsi = \sum_{m_{i}} \psi(m_{1}) \cdots \psi(m_{n})
	          \ket{m_{1}, \dots, m_{n}} \,,
\end{equation}
where the summation is over all possible experimental results.  Since
the experimental setups are all identical, the coefficient function
$\psi(m)$ is the same for all the systems, with
\begin{equation*} \label{eqn:E7}
	\sum_{m} \sqr{\psi(m)} = 1 \,,
\end{equation*}

Each of the states $\ket{m_{1}, \dots, m_{n}}$ in the expansion of
$\ketPsi$ in \eqn{eqn:E6} represents a branch of the universe in which
the specific measurement results $m_{1}, \dots, m_{n}$ are obtained
for the $n$ repetitions of the experiment.  Applying the Born rule, 
as already established for branch states, the probability that the 
universe is in the state $\ket{m_{1}, \dots, m_{n}}$ is 
\begin{equation*} \label{eqn:E7a}
	\mathrm{Prob}(\ket{m_{1}, \dots, m_{n}})
		= \sqr{\psi(m_{1})} \cdots \sqr{\psi(m_{n})} \,.
\end{equation*}
The question to be asked is then how the relative frequency of any
given result $\ell$ in the set $\{m_{i}\}$ compares to the probability
$\sqr{\psi(\ell)}$ predicted by the Born rule \emph{for that
individual result}.

This question has, of course, been addressed before (see, e.g.,
\refcites{hartle68,graham73} for discussions of this question in the
context of traditional and many-worlds interpretations of quantum
mechanics and \refcite{seidewitz06b} for an earlier version of the
argument used below).  However, as noted in \sect{sect:intro}, the use
of relative frequencies to define probability is problematic
\cite{kent90,squires90,zurek05}.  But these problems relate to
attempts to fundamentally justify the Born probability rule itself
using a relative frequency approach.  What we are now addressing is
different: given that the Born probability rule applies for branches
of the state of the universe, we are exploring whether the
\emph{statistics} of repeated measurement results within any such
branch would be expected to follow a similar rule.  In this regard,
criticisms of, e.g., circularity and the need for additional
assumptions, do not apply.

So, we can take the relative frequency for a specific measurement
result $\ell$ within the set $\{m_{i}\}$ to be given by the function
\begin{equation*} \label{eqn:E8}
	f_{\ell}(m_{1},\dots,m_{n}) 
		= \frac{1}{n} \sum_{i=1}^{n} \delta_{m_{i}, \ell} \,.
\end{equation*}
This quantity is itself an observable, for the operator
\begin{equation*} \label{eqn:E9}
	\op{F}_{\ell}\ket{m_{1},\dots,m_{n}}
		= f_{\ell}(m_{1},\dots,m_{n})\ket{m_{1},\dots,m_{n}} \,.
\end{equation*}
The expected value of $\op{F}_{\ell}$ is then
\begin{equation*} \label{eqn:E10}
	\begin{split}
	\langle \op{F}_{\ell} \rangle 
		&\equiv \sum_{m_{1},\dots,m_{n}} f_{\ell}(m_{1},\dots,m_{n})
		        \mathrm{Prob}(\ket{m_{1}, \dots, m_{n}}) \\
		&= \sum_{m_{1},\dots,m_{n}} f_{\ell}(m_{1},\dots,m_{n})
		   \sqr{\psi(m_{1})} \cdots \sqr{\psi(m_{n})} \\
		&= \frac{1}{n} \sum_{m_{1},\dots,m_{n}} \sum_{i=1}^{n} 
		   \delta_{m_{i}, \ell} 
		   \sqr{\psi(m_{1})} \cdots \sqr{\psi(m_{n})} \\
		&= \frac{1}{n} \sum_{i=1}^{n} 
		   \sum_{\substack{m_{1},\dots,m_{i-1}, \\ m_{i+1},\dots,m_{n}}}
		   \sqr{\psi(m_{1})} \cdots \sqr{\psi(m_{i-1})} 
		   \sqr{\psi(\ell)} \\
		&\phantom{= }\qquad\qquad\qquad\qquad\quad
		   \sqr{\psi(m_{i+1})} \cdots \sqr{\psi(m_{n})} \\
		&= \frac{1}{n} \sum_{i=1}^{n} \sqr{\psi(\ell)}
		 = \sqr{\psi(\ell)} \,.
	\end{split}
\end{equation*}

We can now consider the states $\ket{f_{\ell}, m_{1}, \dots, m_{n}}$,
which represent branches of the universe in which a specific relative
frequency is measured for a specific set of experimental results.  The
total probability for measuring a certain $f_{\ell}$ is the sum of the
probabilities for each of the states for which $nf_{\ell}$ of the
$m_{i}$ have the value $\ell$:
\begin{equation*} \label{eqn:E11}
	p(f_{\ell}) = \binom{n}{nf_{\ell}} 
				  \langle \op{F}_{\ell} \rangle^{nf_{\ell}}
	              (1 - \langle \op{F}_{\ell} \rangle)
					  ^{n(1 - f_{\ell})} \,.
\end{equation*}

The probability $p(f_{\ell})$ is a Bernoulli distribution.  By the de
Moivre-Laplace theorem, for large $n$, this distribution is sharply
peaked at the mean $f_{\ell} = \langle \op{F}_{\ell} \rangle =
\sqr{\psi(\ell)}$.  Thus, the probability becomes almost certain that
a choice of one of the states $\ket{f_{\ell}, m_{1}, \dots, m_{n}}$
will be a history in which the observed relative frequency will be
near the prediction given by the usual Born probability
interpretation.  Of course, for finite $n$, there is still the
possibility of a ``maverick'' universe in which $f_{\ell}$ is
arbitrarily far from the expected value---but this is statistically
possible for any probability interpretation over a finite population.
It is not a fundamental problem here, though, because we are not using
relative frequencies to define probability itself, only to interpret
experimental results.

\section{Interpretation} \label{sect:interpretation}

The development of the last four sections can be informally 
summarized as follows:

\begin{enumerate}
	\item \emph{Relativistic quantum mechanics.} Adopt a relativistic 
	formalism in which physical states in Hilbert space represent 
	states of the universe over all spacetime.
	
	\item \emph{Measurement.} Any observable represented as an
	operator on such a state will then induce a decomposition into
	orthogonal eigenstates, each of which represents a potential
	result of measuring the observable.
	
	\item \emph{Probability.} Taking Laplace's ``principle of
	indifference'' as the basis for defining ``probability'', and
	using Zurek's concept of envariance, each such eigenstate has a
	probability of being the actual result that is given by the square
	of the corresponding eigenvalue, that is, by Born's rule.
	
	\item \emph{Statistics} Given Born's rule for the probability of
	eigenstates of the universe, the statistics of experimental
	measurements of an observable will also tend to also follow Born's
	rule (as the number of repeated measurements increases).
\end{enumerate}

The decomposition of a quantum state into decohering branch states is,
of course, common in ``many-worlds'' interpretations of quantum
mechanics.  However, in a timeless relativistic formalism, these
states no longer actually ``branch''.  Instead, each orthogonal
eigenstate in the decomposition of the state of the universe
represents a separate possible history of the entire universe over all
time.  Measurement statistics are seen to be simply the consequence of
an objective probability distribution over this population of
alternative histories.  (While I have not explicitly computed this
distribution, it is essentially the joint probability density of a
``non-commutative model'' given in \refcite{morgan22}.  See also
\refcite{short21}.)

This means that there is no need to make any ontological commitment to
there actually being ``many universes''.  The theory simply predicts
(in principle, for a given Hamiltonian) all the possible decoherent
histories of the universe (with measurements recorded in them) and
assigns a probability to each one that it is the ``real'' history.  It
is then completely consistent for us to assume that there is only one
``real'' universe.  Quantum theory just does not fully determine
what this universe is---we need to do experiments and make
measurements to find out.

And, indeed, as presented here, it is the \emph{recording} of
measurements that actually decoheres the history eigenstates.  Since
each history consists of \emph{records}, it is, by definition, just
the history of what becomes \emph{known} in the universe.  If
something is not measured and recorded, then there is simply no way to
know within the universe whether it happened one way or another.
Thus, recorded measurements effectively result in decoherence, while
lack of measurement allows for interference.

For example, consider the traditional two-slit experiment.  If the
experiment is performed without measuring which slit each particle
goes through, then there will not be separate, decoherent history
eigenstates based on this choice, and the experiment results will
reflect interference.  On the other hand, if the transit of particles
is observed and recorded for even just one of the slits, then there
will be separate history eigenstates based on this observation and no
interference in the experiment results.

Now, despite having a ``timeless'' relativistic formalism, we actually
live our lives moving forward in time.  This means that, at any point
in time, we only have access to records that were made in our past
light cones.  Then, as we move through time, we accumulate new
measurement records.  The so-called ``collapse'' of a quantum state is
simply the result of taking into account the additional knowledge of a
new record gained after carrying out a measurement.

To address this more carefully, note that, in the mathematical
approach developed here, measurements are recorded in the environment
of the measuring apparatus, independently of the system that was
measured.  This means that such records are represented by commuting
observables on the state of the universe.  Therefore, in principle at
least, it is possible to construct joint eigenstates of all the
measurement records that could possibly ever be made in the universe
(consistent with the full state of the universe and the Hamiltonian
constraint representing the physics in it).  These eigenstates form
the complete set of most fine-grained \emph{decoherent histories} of
the entire universe that can be given classical probabilities.

And, obviously, if one parameterizes the state of the universe with
the result of every possible measurement, then the result of every
possible measurement will be completely determined in a universe so
parameterized.  What has been shown in this paper is simply that such
a parameterization can actually be used to label an orthogonal
decomposition of any state of the universe, and that a consistent
Born-rule probability interpretation can be given for the eigenstates
in such a decomposition.

Let $\{\sysS_{i}\}$ be the complete set of systems for which
measurements can be recorded in the universe with state $\ketPsi$.
The decomposition into fine-grained histories is then
\begin{equation*}
	\ketPsi = \sum_{\alpha_{i}}
		\ket{s^{\alpha_{1}}_{1}, s^{\alpha_{2}}_{2}, \dots} \,,
\end{equation*}
where the $\ket{s^{\alpha_{1}}_{1}, s^{\alpha_{2}}_{2}, \dots}$ are
orthogonal eigenstates in which each $\sysS_{i}$ has the recorded
measurement $\alpha_{i}$ (recall that this includes the possibility of
$\alpha_{i} = 0$, meaning that $\sysS_{i}$ does not exist or is not
measured in that history of the universe).

The $s^{\alpha_{i}}_{i}$ completely enumerate the fine-grained history
eigenstates of $\ketpsi$.  In each such eigenstate, each $\sysS_{i}$
is known to be in a specific condition with probability 1.  That is,
\begin{equation*}
	\opP_{\sysS_{i}}^{\beta}
		\ket{s^{\alpha_{1}}_{1}, s^{\alpha_{2}}_{2}, \dots} =
		\delta_{\alpha_{i}\beta}
		\ket{s^{\alpha_{1}}_{1}, s^{\alpha_{2}}_{2}, \dots} \,.
\end{equation*}
If one could truly know all the $s^{\alpha_{i}}_{i}$, then the
eigenstate $\ket{s^{\alpha_{1}}_{1}, s^{\alpha_{2}}_{2}, \dots}$
would, in fact, represent the real history of the universe.  (This is
similar to the idea of ``one real fine-grained history'' proposed by
Gell-Mann and Hartle in a subjective probabilistic context
\cite{gellmann12}.)

Of course, the real history $s^{\alpha_{i}}_{i}$ is not completely
knowable, even in principle.  Not only would it likely require
infinite knowledge, but it would lead to the paradox that, to ``know''
it within the universe, it would itself have to be recorded!  Instead,
all the measurement records that we know of so far cover only a small
subset of the $\sysS_{i}$, determining a vast, coarse-grained
superposition of the fine-grained history eigenstates consistent with
those records:
\begin{equation*}
	\ket{\Psi_{K}} = \sum_{\alpha_{i}}
		\left( \prod_{i \in K} \delta_{\alpha_{i}\kappa_{i}} \right)
		\ket{s^{\alpha_{1}}_{1}, s^{\alpha_{2}}_{2}, \dots} \,,
\end{equation*}
where $K$ is the (relatively small) set of indices of systems with
known measurement records, and the $\kappa_{i}$ are the recorded
measurements for those systems.

Moreover, most of what happens in the expanse of the cosmos has little
or no relevance to the experiments we carry out in our corner of the
universe.  So, the quantum state that we actually use for any specific
system under consideration---${\sysS_{1}}$, say---is reduced over
everything else in the universe:
\begin{equation*}
	\rho_{\sysS_{1}} = \Tr_{\sysS_{i} \neq \sysS_{1}}
		\ket{\Psi_{K}}\bra{\Psi_{K}} \,.
\end{equation*}

Now, suppose $1 \notin K$, and then a measurement of $\sysS_{1}$ is
performed and recorded (obviously involving some other system as the
measurement apparatus, the details of which are not important here).
The result is that we now, in principle, know the state of the
universe a little more precisely, as $\ket{\psi_{K'}}$, where $K' = K
\cup \{1\}$ and $\sysS_{1}$ has the known condition $\kappa_{1}$.  But
$\ket{\psi_{K'}}$ is then an eigenstate of
$\opP_{\sysS_{1}}^{\kappa_{1}}$ and the corresponding reduced state
$\rho'_{\sysS_{1}}$ is similarly an eigenstate for the reduced state
space of $\sysS_{1}$, with eigenvalue $\kappa_{1}$.

This is the effective ``collapse'' of the state of $\sysS_{1}$ after a
measurement.  The update of the local reduced state for $\sysS_{1}$
seems discontinuous, but that is only because we have ignored the
implicit parts of the earlier state that integrated over the possible
choices of the future measurement.  An update exactly projects the
original state into a more constrained Hilbert subspace consistent
with the new measurement result, and the new state is then a bit
``closer'' to the actual ``real'' history of the universe.

Essentially, this can be considered a relativistic generalization of a
consistent histories interpretation of quantum mechanics
\cite{griffiths84,griffiths02}.  However, instead of a ``history''
being a sequence of measurement projections made over time, a history
is a set of projections representing conditions of the universe across
spacetime \cite{seidewitz06b,seidewitz11}.  Of course, it is still
possible to choose observables that represent conditions at specific
points in time, and then organize them in a time-ordered history, in
which case the relativistic generalization reduces to the traditional
non-relativistic approach, as would be expected.

\section{Conclusion} \label{sect:conclusion}

In \refcite{adlam23}, Adlam proposes that any viable solution to the 
measurement problem will have the following four features:

\begin{enumerate}
	\item  ``It makes no substantial change to the formalism of unitary 
	quantum mechanics (at least at the microscopic level).''

	\item  ``Decoherence plays a significant role in the emergence of 
	classical reality.''

	\item  ``Observers (approximately) see a unique outcome to each 
	measurement and are able to (approximately) establish a shared 
	observable reality.''

	\item  ``This shared observable reality supervenes on beables which 
	are approximate and emergent, and/or non-dynamical, and/or 
	non-microscopically defined.''
\end{enumerate}

The approach I have proposed in this paper satisfies these criteria.
It requires no change at all to the formalism of unitary RQM, and it
takes a modern Everettian view of decoherence.  As discussed in
\sect{sect:interpretation}, observers are expected to see unique,
objective outcomes to measurement processes, and the recorded results
of these measurements are, in fact, the ``beables'' of the theory.
Finally, the timeless 4-dimensional spacetime formalism I have used
makes the proposal a ``non-dynamical'' approach (in Adlam's terms).

Indeed, as ``beables'', one can consider the fine-grained measurement
records $s^{\alpha_{i}}_{i}$ (defined in \sect{sect:interpretation})
to act much like ``hidden variables'', since they completely determine
the results of every measurement made in the universe with eigenstate
$\ket{s^{\alpha_{1}}_{1}, s^{\alpha_{2}}_{2}, \dots}$.  However, they
are truly ``hidden'', since each value $\alpha_{i}$ can literally only
be determined within the universe by actually making a measurement and
recording that value.

So, on the one hand, such states $\ket{s^{\alpha_{1}}_{1},
s^{\alpha_{2}}_{2}, \dots}$ can be considered an \emph{ontic}
representation of the real history of the universe.  On the other
hand, quantum states as we typically use them are pragmatically
\emph{epistemic}.  They record our knowledge on some small parts of
the universe, based on the measurements that we have recorded so far,
and they get updated as we get new knowledge.

Now, a local reduced state $\rho_{\sysS}$ (also as defined in
\sect{sect:interpretation}) can actually be deterministically and
uniquely computed from the underlying $\ket{s^{\alpha_{1}}_{1},
s^{\alpha_{2}}_{2}, \dots}$ states.  However, the point is that we
cannot ever really know, even in principle, \emph{how} to carry out
this computation without actually making measurements of the system
$\sysS$.  As a result we are forced to consider $\rho_{\sysS}$ as
simply representing our ``best knowledge'' of $\sysS$, defining a
probability distribution for the result of measuring the system, over
the underlying ``hidden'' fine-grained states of the universe.

(It is worth noting that I am using the terms ``ontic'' and
``epistemic'' here in the traditional philosophical sense, \emph{not}
in the technical sense of ``$\psi$-ontic'' and ``$\psi$-epistemic'' as
defined by Harrigan and Spekkens \cite{harrigan10}.  Indeed, by
Harrigan and Spekkens' definition, the local reduced state
$\rho_{\sysS_{1}}$ is $\psi$-ontic, while, as I have argued, in any
pragmatic sense, it is epistemic.  See also \refcite{hance22a}.)

This viewpoint on states such as $\rho_{\sysS}$ is similar to the
$\psi$-ensemble interpretation of \refcite{hance22b}, in which the
wave function is interpreted as representing a true ensemble over
hidden variable states.  Nevertheless, the quantum formalism used here
is entirely orthodox, so, while the interpretation explains what
``collapse'' means, it sheds little further light on other
particularly quantum effects such as interference.
Interpretationally, this may be the best we can do without truly
introducing additional ``hidden variables'' into the formalism.